\documentclass[a4paper,11pt]{article}
\usepackage{pos}

\title{Jet charge modification in dense QCD matter}

\author[a]{Haitao Li}
\author*[b]{Ivan Vitev}

\affiliation[a,b]{Los Alamos National Laboratory,\\
 Mail Stop B283, Los Alamos, NM 87545, USA}

\emailAdd{haitaoli@lanl.gov}
\emailAdd{ivitev@lanl.gov}

\abstract{In these proceedings we report a recent calculation of the jet charge modification in heavy-ion relative to proton collisions at the LHC.   Jets have played an essential role in constraining  theories of  in-medium parton shower evolution and in determining the properties of the quark-gluon plasma created in  ultra-relativistic nuclear  reactions.  It is important  to extend these studies to flavor-tagged jets and explore observables that are sensitive to their partonic origin. The average jet charge, introduced early on in the history of quantum chromodynamics, is a proxy for the electric charge of the quark or gluon that initiates the jet.  In the framework of soft-collinear effective theory, we show how  to  evaluate the jet charge in a dense strongly-interacting matter environments. We identify observables that can isolate the contribution of in-medium branching from  isospin effects and present predictions for the transverse momentum dependence of the jet charge distribution in nucleus-nucleus collisions and its modification relative to the proton case.}

\FullConference{%
  HardProbes2020\\
  1-6 June 2020\\
  Austin, Texas}

\begin{document}
\maketitle

\section{Introduction}

The jet charge is a substructure observable designed to approximate the electric charge of the hard scattered parton that initiates the jet. It was introduced in the late 1970s~\cite{Field:1977fa}   and is defined as the transverse momentum-weighted sum of the charges of particles within the jet cone
\begin{align} \label{eq:charge}
    Q_{\kappa, {\rm jet}}  = \frac{1}{\left(p_T^{\rm jet}\right)^\kappa } \sum_{\rm i\in jet} Q_i \left(p_T^{i} \right)^{\kappa } \;  .
\end{align}
Here, $Q_i$ and $p_T^{i}$ are the electric charge  and the transverse momentum of  particle $i$, and  $\kappa > 0$ is a free parameter. From the point of view of heavy-ion
physics, the ability to identify the partonic origin of jets is extremely useful, as the modification in nuclear matter is different for quark and gluon 
jets~\cite{Chien:2015hda}.  Jet charge  calculations for lead-lead (Pb+Pb) collisions at the LHC have been performed using a monte carlo approach~\cite{Chen:2019gqo}  and the framework of soft-collinear effective theory (SCET)~\cite{Li:2019dre}. In these proceedings we review the latter. First measurements of the jet charge in heavy-ion collisions have also appeared and have been used to isolate the fraction of gluon-like jets~\cite{Sirunyan:2020qvi}.

Starting with the definition Eq.~(\ref{eq:charge}) and realizing that gluons do not contribute to the average jet charge, this observable can be expressed as follows:
 \begin{align}
    \langle Q_{\kappa, q} \rangle = \int dz~z^\kappa \sum_{h} Q_h \frac{1}{\sigma_{\text{q-jet}}} \frac{d\sigma_{h\in \text{q-jet}}}{dz} \;, \quad
    \langle Q_{\kappa, q} \rangle = \frac{\tilde{\mathcal{J}}_{qq}(E,R,\kappa,\mu)}{J_{q}(E,R,\mu)} \tilde{D}_q^{Q}(\kappa,\mu) \;,
\end{align}
where here $J_{q}(E,R,\mu)$ is a jet function. $\tilde{\mathcal{J}}_{qq}(E,R,\kappa,\mu)$ is the Wilson coefficient for  matching  the quark  fragmenting  jet  function  onto  a quark fragmentation  function and  $\tilde{D}_q^{Q}(\kappa,\mu)$ is a fragmentation function~\cite{Krohn:2012fg}.  
The $(\kappa+1)$-th Mellin moments of the jet matching coefficient and fragmentation function are defined as
\begin{align} \label{eq:moments}
    \tilde{\mathcal{J}}_{qq}(E,R,\kappa,\mu) &= \int_0^1 dz~ z^{\kappa} \mathcal{J}_{qq}(E,R,z,\mu)  \;,  \quad
    \tilde{D}_q^{Q}(\kappa,\mu) = \int_0^1 dz ~z^{\kappa} \sum_{h} Q_h D_q^{h}(z,\mu) \;.
\end{align}
In Eq.~(\ref{eq:moments}) $z= p_T^{i}/p_T$, $E$ is the jet energy,  $R$ is the jet radius, and $\mu$ is the factorization scale. An important property of the jet charge is
that it is sensitive to scaling violations in QCD 
\begin{align} \label{eq:scalev}
    \frac{p_T}{\langle Q_{\kappa, q}  \rangle} \frac{d}{dp_T}\langle Q_{\kappa, q} \rangle = \frac{\alpha_s}{\pi} \tilde{P}_{qq}(\kappa) \; ,
\end{align}
where $ \tilde{P}_{qq}(\kappa)$ is the $(\kappa+1)$-th Mellin moment of the leading order splitting function.  The effect has been measured in 
proton-proton collisions~\cite{Aad:2015cua} and this serves as a strong motivation to extend the
observable to heavy-ion collisions. 

\section{Theoretical formalism in heavy ion collisions  and numerical results}

\begin{figure}[t!]
    \centering
        \includegraphics[scale=0.37]{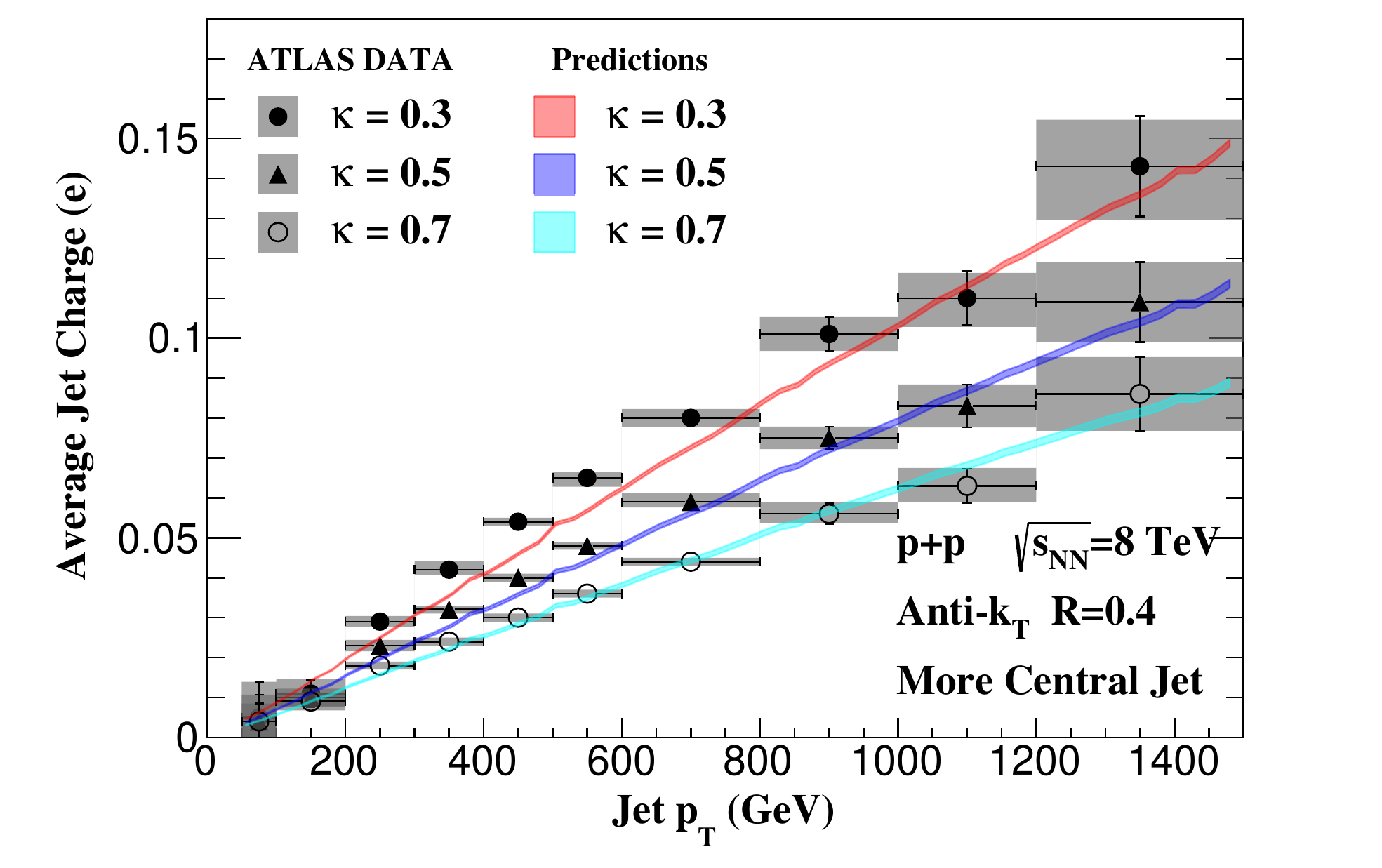}    \hspace*{0.05in}
    \includegraphics[scale=0.35]{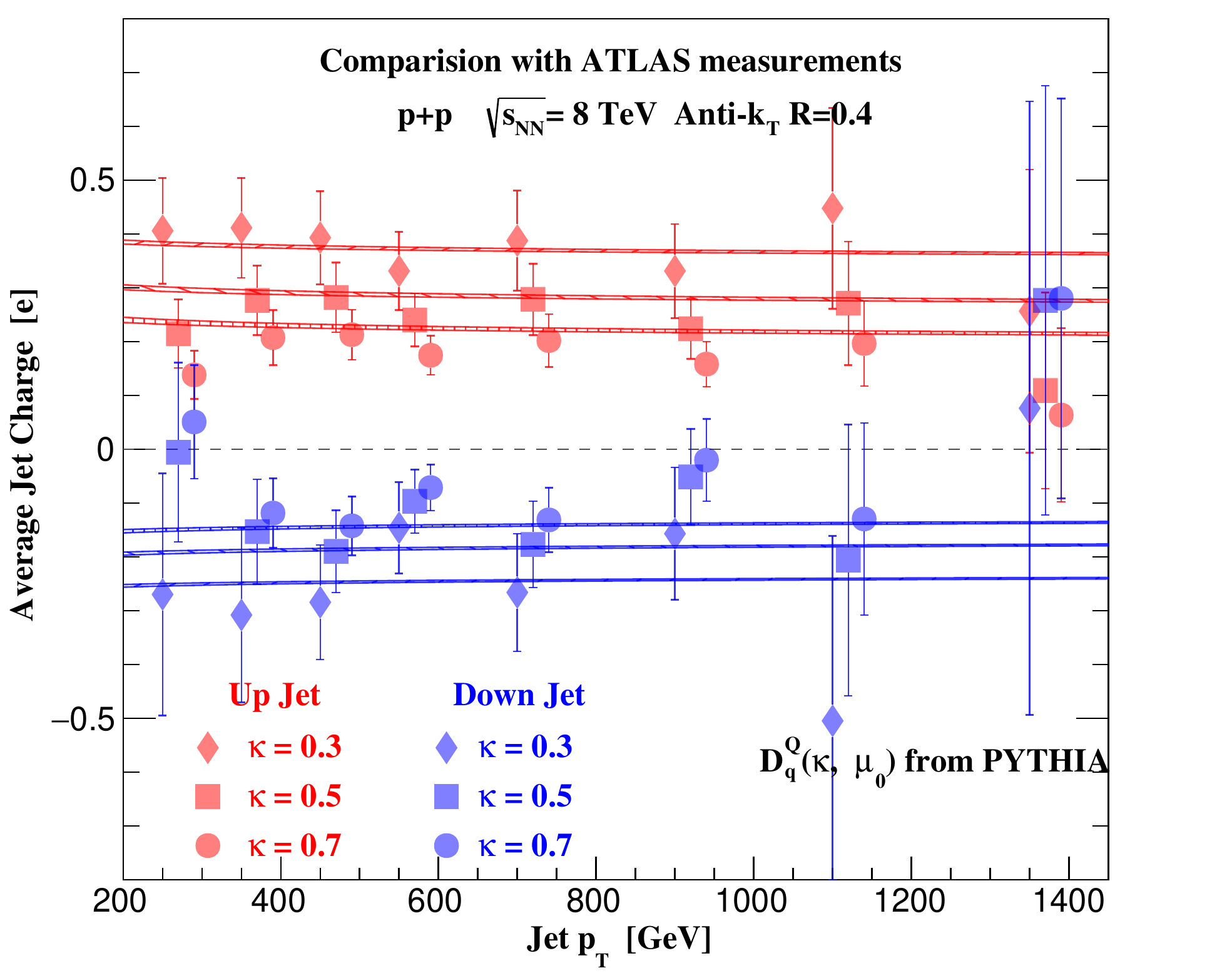}
    \vspace{-0.2cm}
    \caption{Left:  transverse momentum dependence of the average  jet charge distribution with  $\kappa=$0.3, 0.5 and 0.7 for the more central jets  in $\sqrt{s_{\rm NN}} = 8$~TeV  p+p collisions at the LHC.  Right: average charge of up and down-quark jets as a function of jet $p_T$. Data is from ATLAS~\cite{Aad:2015cua}.  }
    \label{fig:ppQ}
\end{figure}

Before we proceed to the evaluation of the jet charge in Pb+Pb collisions we will validate the SCET formalism in the simpler p+p reactions. The ATLAS collaboration has performed measurements of back-to-back jets at $\sqrt{s}=8$~TeV, denoting them as a more forward and a more central jet, and extracted the flavor dependent jet charge 
\begin{align}\label{eq:Qjet}
    \langle Q^{f/c}_{\kappa} \rangle = (f^{f/c}_{u}-f^{f/c}_{\bar{u}})  \langle Q_{\kappa}^{ u} \rangle+(f^{f/c}_{d}-f^{f/c}_{\bar{d}})  \langle Q_{\kappa}^{ d} \rangle \, .
\end{align}
In Eq.~(\ref{eq:Qjet}) $f^{f/c}_q$ is the fraction   of $q$-flavored jets for the more forward/central jets and $  \langle Q_{\kappa}^{ q} \rangle $ is the average charge for the $q$ jet.
Our theoretical results for the average jet charge and  the up- and down-quark jet charges as a function of jet $p_T$ are shown in Fig.~\ref{fig:ppQ}.
The average jet charge only relies on one non-perturbative parameter/boundary condition for a given $\kappa$ and the jet type, which we obtain through PYTHIA simulations.  The uncertainties are evaluated by varying the factorization scale $\mu$ by a factor of two.  The left panel of  Fig.~\ref{fig:ppQ} gives the average jet charge for more central jets  and its absolute value decreases with $\kappa$, as expected from  Eq.~(\ref{eq:charge}).  The right  panel of  Fig.~\ref{fig:ppQ} gives the flavor-separated  
charges  for up- and down-quark jets. The predictions agree very well with the measurements by ATLAS~\cite{Aad:2015cua}, even though the data have large experimental uncertainties.

Propagation of partons in QCD matter adds a medium-induced component to the parton showers that characterize simpler reactions. The in-medium branching processes relevant to shower formation can be calculated  order-by-order in powers of the mean number of scatterings~\cite{Sievert:2019cwq}. An important characteristic of medium-induced showers, which persists to higher orders in $\alpha_s$ \cite{Fickinger:2013xwa},  is that they are softer and broader than the vacuum ones. Jet production and jet substructure in reactions with nuclei can be evaluated in a systematic and improvable fashion using a generalization of SCET to include interactions between its degrees of freedom and QCD matter mediated by Glauber gluons (SCET$_{\rm G}$). Thus, the ingredients of SCET factorization receive medium corrections where relevant. For example,  QGP contribution to the the matching coefficients  can be expressed in terms of the in-medium splitting kernels
\begin{align}
    \mathcal{J}_{qq,(qg)}^{\rm med }(E,R,x,\mu)
    =\frac{\alpha_s(\mu)}{2\pi^2}  \int_0^{2 E x(1-x)\tan R/2} \frac{d^2 \mathbf{k}_{\perp} }{\mathbf{k}_{\perp}^2}  P_{q \rightarrow q g,(gq)}^{\rm{med}}\left(x, \mathbf{k}_{\perp}\right) \,.
\label{match}
\end{align}
 The medium correction to the full quark jet function reads
\begin{align}
    J_{q}^{\rm med}(E,R,\mu)  &= 
    \int_0^1 dx~ x \bigg(\mathcal{J}_{qq}^{\rm med}(E,R,x,\mu)+\mathcal{J}_{qg}^{\rm med}(E,R,x,\mu)\bigg)   \\
  =  &\frac{\alpha_s(\mu)}{2\pi^2} \int_0^1 dx  \int_{0}^{2 E x(1-x)\tan R/2} \frac{d^2 \mathbf{k}_{\perp} }{\mathbf{k}_{\perp}^2 }
    P_{q \rightarrow  qg }^{\rm{med,real}}\left(x, \mathbf{k}_{\perp} \right),
 \label{inclfunc}
\end{align} 
see also \cite{Kang:2017frl}. Finally,  in a QCD medium the evolution of the charge-weighted fragmentation function becomes
\begin{align} \label{eq:AAQevol}
  \frac{d}{d\ln \mu} \tilde{D}_q^{Q, {\rm full}}(\kappa,\mu)  =
   \frac{\alpha_s(\mu)}{\pi} \left(\tilde{P}_{qq}(\kappa)+\tilde{P}^{\rm med}_{qq}(\kappa,\mu) \right) \tilde{D}_q^{Q,{\rm full}}(\kappa,\mu) \; ,
\end{align}
where $\tilde{P}^{\rm med}_{qq}(\kappa,\mu)$ is the $(\kappa+1)$-th Mellin moment of the medium splitting kernel. 
The additional scale dependence in the medium-induced part of Eq.~(\ref{eq:AAQevol}) reflects the difference in the $k_\perp$ dependence of the vacuum and in-medium 
 branching processes~\cite{Kang:2014xsa}.

 \begin{figure}[t!]
    \centering
        \includegraphics[scale=0.36]{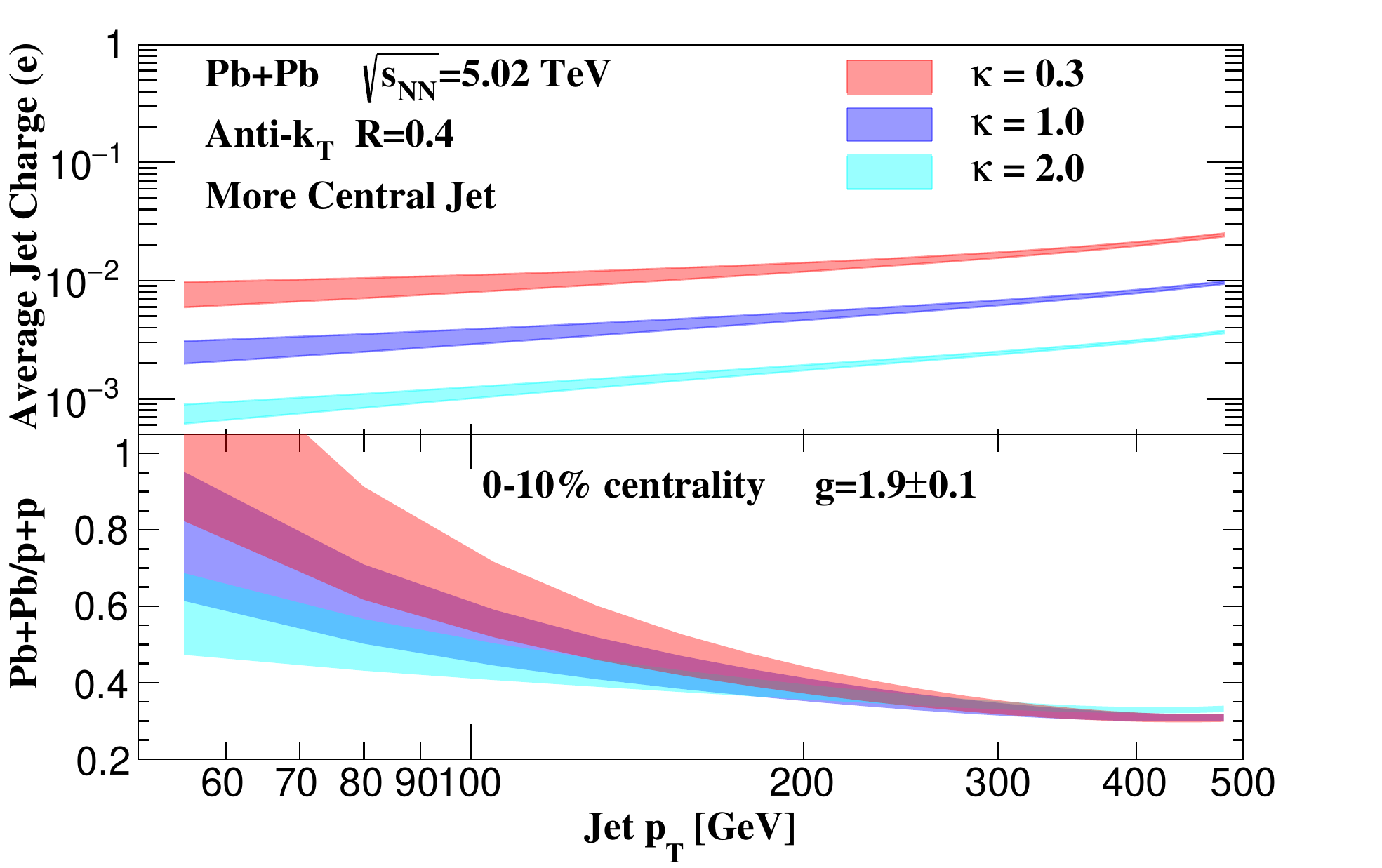}    \hspace*{-0.2in}
    \includegraphics[scale=0.41]{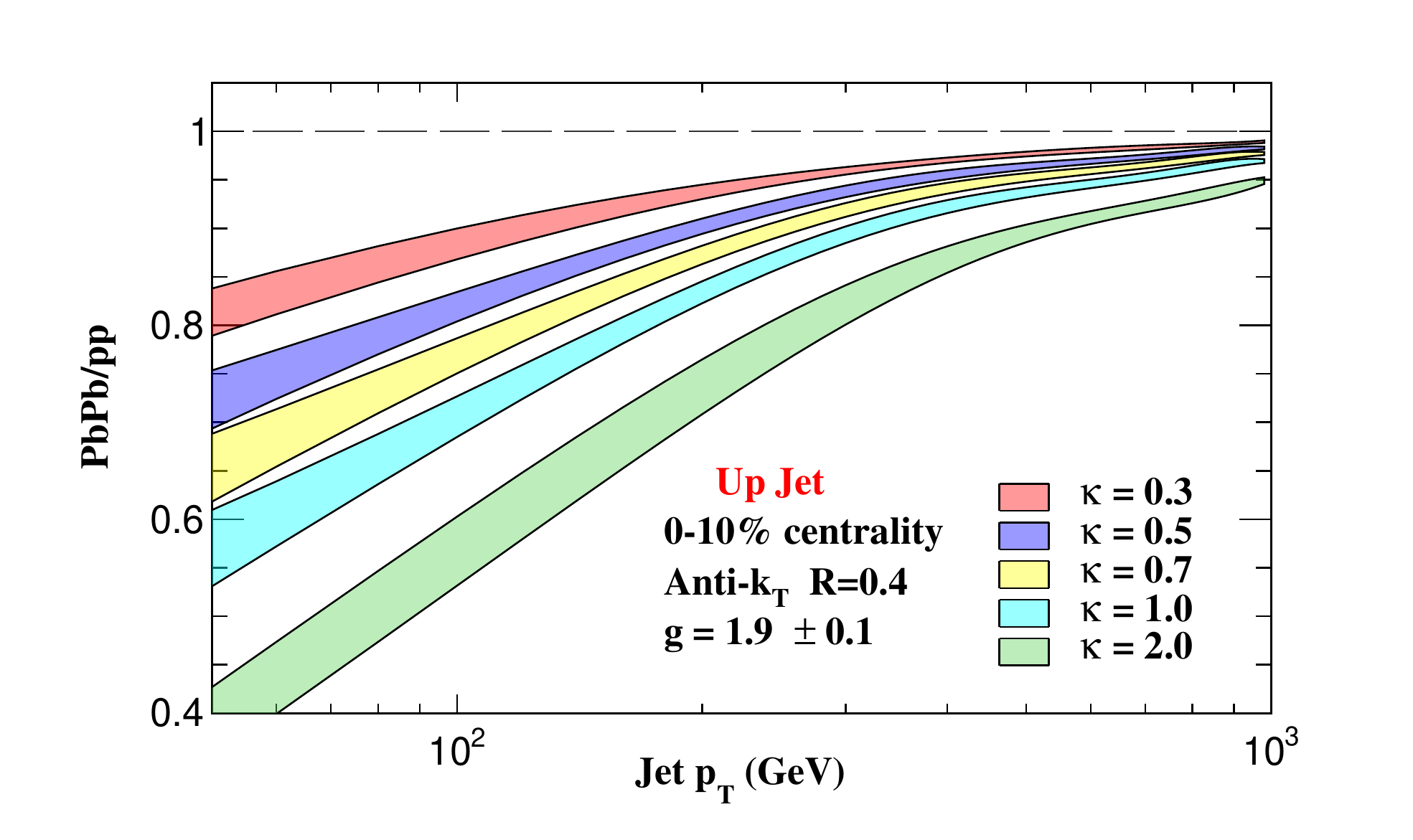}
    \vspace{-0.cm}
    \caption{Left:  The average jet charge in  $\sqrt{s_{\rm NN}} = 5.02$~TeV central Pb+Pb collisions for more central jets and its modification relative to p+p collisions.  Calculations for  $\kappa=$0.3, 1 and 2  are shown.    Right: Modification of the up-jet charge due to in-medium evolution as a function of transverse momentum.      }
    \label{fig:AAQ}
\end{figure}

 The jet charge and  its modification in central Pb+Pb collisions at the LHC are shown in  the left panel of  Fig.~\ref{fig:AAQ}.  At very high transverse momenta it is completely dominated by isospin effects. However, for $p_T < 200$~GeV  one begins to observe the effects of in-medium evolution. The uncertainty bands correspond to the variation of the  coupling $g$ between the jet and the medium in the interval (1.8,2.0). The need to cleanly isolate the contribution of in-medium evolution to jet charge modification led us to propose a new observable  -- 
the modification of individual flavor jet charge in heavy-ion versus proton collisions.   This can be seen  in  the right panel of  Fig.~\ref{fig:AAQ}
where we show the medium modifications to the up-quark jet charge.  The only difference between the up- and down-quark jet charges is the fragmentation function boundary condition, hence their modification is the same
\begin{align}
    \frac{\langle Q_{\kappa, u}^{ \rm Pb+Pb}(p_T)\rangle }{\langle Q_{\kappa, u}^{\rm p+p}(p_T)\rangle} =\frac{\langle Q_{\kappa,d}^{ \rm Pb+Pb}(p_T)\rangle }{\langle Q_{\kappa, d}^{\rm p+p}(p_T)\rangle} \; . 
\end{align} 
The individual jet charge modification eliminates the initial-state isospin effects and helps reveal the  final-state medium-induced parton shower  contribution to  the
jet function and the fragmentation function evolution.  For this reason, the medium corrections are larger for  smaller energy jets - a kinematic region where the medium-induced splitting functions are more important. Furthermore, when $\kappa$ is large the $(\kappa+1)$-th Mellin moment of the medium splitting function is more sensitive to  soft-gluon emission.

\section{Conclusions}

We  presented recent calculation of the jet charge distributions in heavy-ion collisions in the SCET$_{\rm G}$ effective field theory framework~\cite{Li:2019dre}.  
In the presence of nuclear matter the jet functions, jet matching coefficients, and the evolution of the fragmentation functions are constructed with the help of the medium-induced splitting kernels. 
The jet charge observable is particularly interesting because of its  ability to discriminate between jets of various flavors, for example  up-quark jets, down-quark jets  and gluon jets.
This discriminating power remains valid in nucleus-nucleus collisions. Furthermore,  the charge of  jets  can provide novel insight into the Mellin moments of medium-induced splitting functions and the in-medium evolution of the  non-perturbative fragmentation functions. 

The jet charge definition is independent of the hard process. Thus, jet charge modification can be studied in other types of nuclear matter such as e+A collisions at the future electron-ion collider (EIC).   Recent calculations of light and heavy meson production at the EIC  have shown that with appropriate choice of center-of-mass energies and rapidity domains jet quenching effects in cold nuclear matter can be large and observable~\cite{Li:2020zbk}.  We plan to evaluate the jet charge in e+A reactions in the future.


\begin{thebibliography}{99}

\bibitem{Field:1977fa}
R.~D.~Field and R.~P.~Feynman,
Nucl. Phys. B \textbf{136}, 1 (1978)


\bibitem{Chien:2015hda}
Y.~T.~Chien and I.~Vitev,
JHEP \textbf{05}, 023 (2016)
[arXiv:1509.07257 [hep-ph]].


\bibitem{Chen:2019gqo}
S.~Y.~Chen, B.~W.~Zhang and E.~K.~Wang,
Chin. Phys. C \textbf{44}, no.2, 024103 (2020)
[arXiv:1908.01518 [nucl-th]].


\bibitem{Li:2019dre}
H.~T.~Li and I.~Vitev,
Phys. Rev. D \textbf{101}, 076020 (2020)
[arXiv:1908.06979 [hep-ph]].

\bibitem{Sirunyan:2020qvi}
A.~M.~Sirunyan \textit{et al.} [CMS],
JHEP \textbf{07}, 115 (2020)
[arXiv:2004.00602 [hep-ex]].


\bibitem{Krohn:2012fg}
D.~Krohn, M.~D.~Schwartz, T.~Lin and W.~J.~Waalewijn,
Phys. Rev. Lett. \textbf{110}, no.21, 212001 (2013)
[arXiv:1209.2421 [hep-ph]].


\bibitem{Aad:2015cua}
G.~Aad \textit{et al.} [ATLAS],
Phys. Rev. D \textbf{93}, no.5, 052003 (2016)
doi:10.1103/PhysRevD.93.052003
[arXiv:1509.05190 [hep-ex]].

\bibitem{Fickinger:2013xwa}
M.~Fickinger, G.~Ovanesyan and I.~Vitev,
JHEP \textbf{07}, 059 (2013)
[arXiv:1304.3497 [hep-ph]].

\bibitem{Sievert:2019cwq}
M.~D.~Sievert, I.~Vitev and B.~Yoon,
Phys. Lett. B \textbf{795}, 502-510 (2019)
[arXiv:1903.06170 [hep-ph]].

\bibitem{Kang:2017frl}
Z.~B.~Kang, F.~Ringer and I.~Vitev,
Phys. Lett. B \textbf{769}, 242-248 (2017)
[arXiv:1701.05839 [hep-ph]].

\bibitem{Kang:2014xsa}
Z.~B.~Kang, R.~Lashof-Regas, G.~Ovanesyan, P.~Saad and I.~Vitev,
Phys. Rev. Lett. \textbf{114}, no.9, 092002 (2015)
[arXiv:1405.2612 [hep-ph]].

\bibitem{Li:2020zbk}
H.~T.~Li, Z.~L.~Liu and I.~Vitev,
[arXiv:2007.10994 [hep-ph]].

\end{thebibliography}
\end{document}